\documentclass[12pt]{article}

\usepackage{amsfonts}
\usepackage{amsmath}
\usepackage{amsbsy}
\usepackage{graphicx}
\usepackage{amssymb}
\usepackage{amsthm}
\usepackage{bm}
\usepackage{epsfig}
\usepackage{latexsym}
\usepackage{pdflscape}
\usepackage{color}
\numberwithin{equation}{section}
\input amssym.def
\input amssym.tex
\allowdisplaybreaks

\DeclareMathOperator{\Ai}{Ai}
\DeclareMathOperator{\vol}{vol}

\def\Jt{{\widehat J}}
\def\Jn{J}

\newcounter{aff}

\begin{document}

\begin{titlepage}
\begin{flushright}
{\footnotesize HRI/ST1403, KEK-TH-1714}
\end{flushright}
\begin{center}
{\Large\bf Instanton Effects in Orbifold ABJM Theory}

\bigskip\bigskip
{\large Masazumi Honda\footnote[1]{\tt masazumihonda@hri.res.in}
\quad and \quad
Sanefumi Moriyama\footnote[2]{\tt moriyama@math.nagoya-u.ac.jp}
}\\
\bigskip
${}^{*}$\,
{\small\it Harish-Chandra Research Institute\\
Chhatnag Road, Jhusi, Allahabad 211019, India,}
\medskip\\
{\small\it High Energy Accelerator Research Organization (KEK)\\
Tsukuba, Ibaraki 305-0801, Japan}
\bigskip\\
${}^{\dagger}$\,
{\small\it Graduate School of Mathematics, Nagoya University\\
Nagoya 464-8602, Japan,} 
\medskip\\
{\small\it Kobayashi Maskawa Institute, Nagoya University\\
Nagoya 464-8602, Japan,} 
\medskip\\
{\small\it Yukawa Institute for Theoretical Physics, Kyoto University\\
Kyoto 606-8502, Japan}
\end{center}

\begin{abstract}
We study the partition function of the orbifold ABJM theory on $S^3$,
which is the $\mathcal{N}=4$ necklace quiver Chern-Simons-matter
theory with alternating levels, in the Fermi gas formalism.
We find that the grand potential of the orbifold ABJM theory is
expressed explicitly in terms of that of the ABJM theory.
As shown previously, the ABJM grand potential consists of the naive
but primary non-oscillatory term and the subsidiary
infinitely-replicated oscillatory terms.
We find that the subsidiary oscillatory terms of the ABJM theory
actually give a non-oscillatory primary term of the orbifold ABJM
theory.
Also, interestingly, the perturbative part in the ABJM theory results
in a novel instanton contribution in the orbifold theory.
We also present a physical interpretation for the non-perturbative
instanton effects.
\end{abstract}

\begin{center}
\includegraphics[scale=0.3]{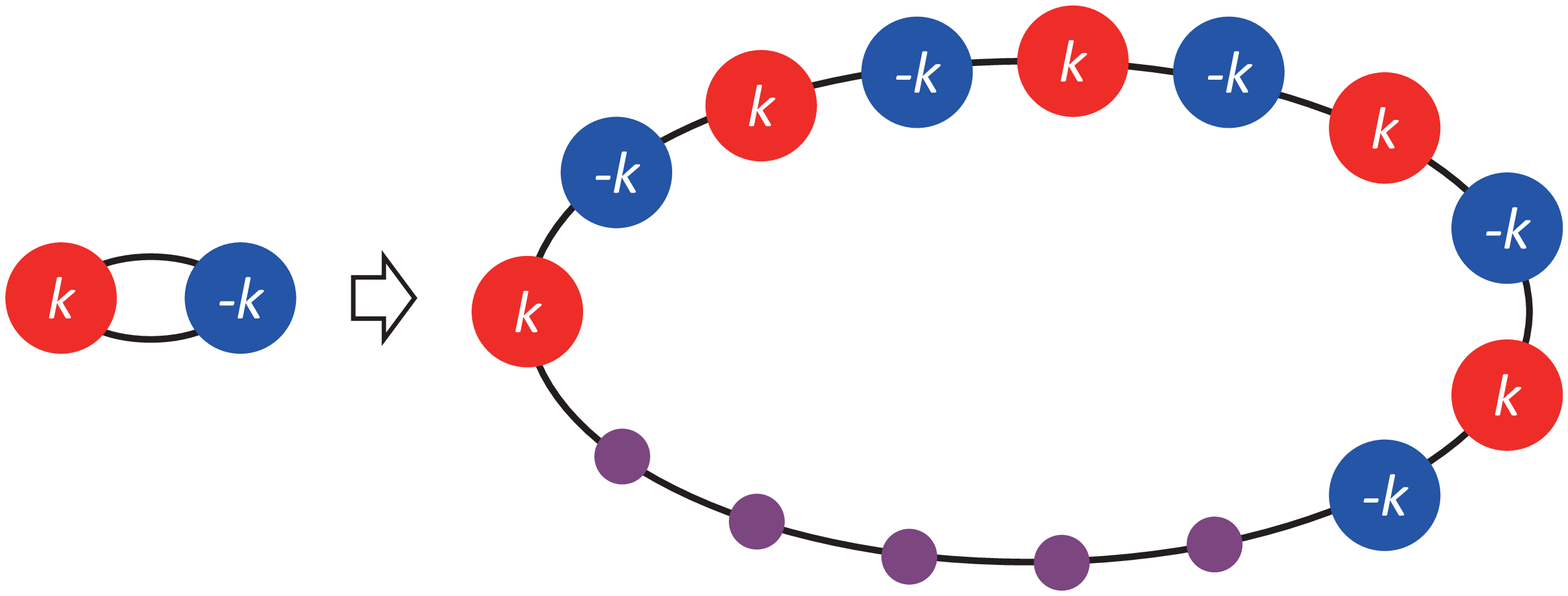}
\end{center}

\end{titlepage}
\tableofcontents

\section{Introduction and Summary}\label{intro}

\begin{figure}[tb]
\begin{center}
\includegraphics[scale=0.4]{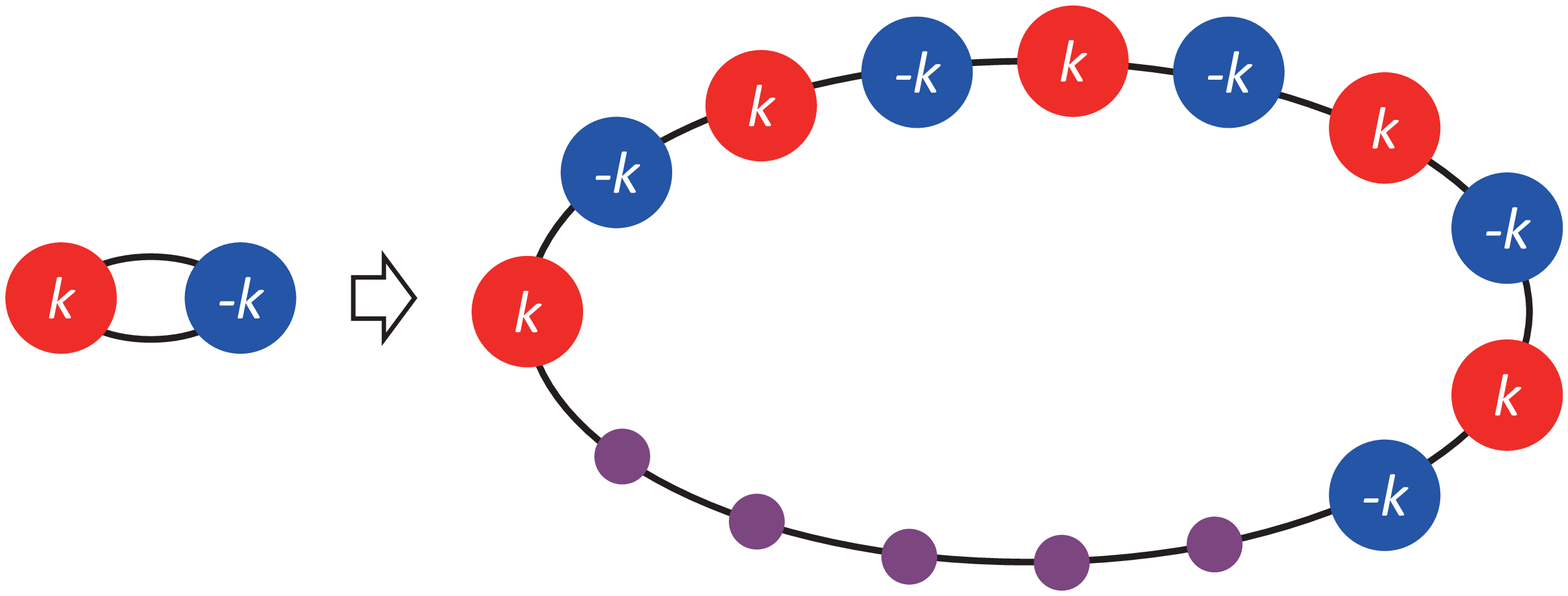}\\
\hspace{-2cm}(a)\hspace{6cm}(b)
\caption{The quiver diagrams of the ABJM theory (a) and the orbifold
ABJM theory (b).
The orbifold ABJM theory is the Chern-Simons-matter theory with gauge
group $[U(N)\times U(N)]^r$ and alternating levels $k$ and $-k$.
Each node represents the $U(N)$ vector multiplet, while each line
represents the bi-fundamental hypermultiplet.
}
\label{fig:quiver}
\end{center}
\end{figure}

Recently there has been a breakthrough in understanding M2-branes in
M-theory.
It was found by Aharony, Bergman, Jafferis and Maldacena \cite{ABJM}
that the worldvolume theory of $N$ multiple M2-branes on
${\mathbb C}^4/{\mathbb Z}_k$ is described by the ${\mathcal N}=6$
Chern-Simons-matter theory with gauge group $U(N)\times U(N)$ and
levels $k$ and $-k$.
Prior to this important discovery, the program of finding the
worldvolume theory of multiple M2-branes by supersymmetrizing the
three-dimensional Chern-Simons theory dates back to the pioneering
studies in \cite{S}.
Up to ${\mathcal N}=3$, supersymmetric Chern-Simons-matter theories
were constructed for any gauge group and any representation
\cite{ZK,KL,KLL}.
For ${\mathcal N}=4$, one of the interesting realizations is the
quiver gauge theory with gauge group being $[U(N)\times U(N)]^{r}$
$(r\in{\mathbb N})$ and levels $k$ and $-k$ appearing alternatively
\cite{GW,HLLLP1}.
Especially, if we consider the case of $r=1$, the supersymmetry is
enhanced to ${\mathcal N}=6$ and this is nothing but the ABJM theory.
The quiver diagram of the ABJM theory is the Dynkin diagram of the
affine Lie algebra $\widehat A_1$, while that of the $\mathcal{N}=4$
theory is the Dynkin diagram of $\widehat A_{2r-1}$ (see figure
\ref{fig:quiver}).
Also, the gravity dual of the ABJM theory is
$AdS_4\times S^7/{\mathbb Z}_k$, while that of the $\mathcal{N}=4$
theory is $AdS_4\times S^7/({\mathbb Z}_r\times{\mathbb Z}_{kr})$.
Since the gravity dual of this theory was identified to be the
orbifold of the ABJM theory \cite{BKKS,IK,TY,IK4}, let us call this
theory orbifold ABJM theory.

For the ABJM case, the partition function and the vacuum expectation
values of the supersymmetric Wilson loops were extensively studied in
\cite{DT,MPtop,DMP1,HKPT,DMP2,FHM,O,MP,KEK,KMSS,HMO1,PY,HMO2,CM,HMO3,
GKM,HMMO,HHMO,KM}.
First, using the localization technique \cite{P}, the
infinite-dimensional path integral used in defining the expectation
values of supersymmetric quantities in general $\mathcal{N}\geq 2$
supersymmetric gauge theories on $S^3$ is reduced to a
finite-dimensional matrix integral \cite{KWY,DT,MPtop,J,HHL}.
After the standard matrix model analyses in the 't Hooft limit
\cite{DT,MPtop,DMP1,HKPT,DMP2,FHM}, the partition function for a
general $\mathcal{N}\geq 3$ necklace quiver Chern-Simons-matter theory
was rewritten into that of an ideal Fermi gas system \cite{MP}, which
is more suitable to access the M-theory regime (see also
\cite{O,KWYmirror}).
This formalism further enables us to continue to study instanton
effects in the ABJM theory \cite{HMO1,PY,HMO2,CM,HMO3,HMMO,HHMO,KM},
where an infinite cancellation of divergences between worldsheet
instantons and membrane instantons was found \cite{HMO2}.
Finally it turned out \cite{HMMO} that the instanton effects in the
ABJM theory are described by certain limits of the refined topological
string on the dual geometry, local $\mathbb{P}^1\times\mathbb{P}^1$.

The aim of this paper is to study how the interesting instanton
calculus of the supersymmetric quantities in the ABJM theory is
generalized to a larger class of theories.
Particularly, as a first step, we shall study the partition function
of the ${\mathcal N}=4$ orbifold ABJM theory on $S^3$, which is
expected to be the simplest one compared with other ${\mathcal N}=3$
Chern-Simons-matter theories.
For the studies in the M-theory limit, of other gauge groups or other
quiver diagrams which are expected to have interesting M-theory
interpretations, see
e.g.~\cite{MS,CHH,JKPS,GHP,GAH,GHN,interacting,MePu,GM} and see
\cite{SMP,Suyama1,Suyama2,Suyama3} for those in the 't Hooft limit.

Before explaining our work, let us briefly overview the ABJM partition
function.
After the study of the large $N$ behavior in the 't Hooft limit in a
seminal paper \cite{DMP1}, it was found in \cite{FHM} that all the
perturbative corrections are summed up into the Airy function
\begin{align}
Z_1(N)\sim C^{-1/3}\Ai\bigl[C^{-1/3}N\bigr]
=\int_{-\infty i}^{\infty i}\frac{d\mu}{2\pi i}
e^{\frac{C}{3}\mu^3-\mu N},
\label{airy}
\end{align}
with a coefficient $C$.
This integral expression is reminiscent of the statistical mechanics.
Namely, if we consider the grand potential
\begin{align}
e^{\Jt_1(\mu)}=\sum_{N=0}^\infty e^{\mu N}Z_1(N),
\label{J1def}
\end{align}
by regarding $N$, the rank of the gauge group, as the number of
particles and introducing a chemical potential $\mu$, we find that the
inverse transformation given by
\begin{align}
Z_1(N)=\int_{-\pi i}^{\pi i}\frac{d\mu}{2\pi i}e^{\Jt_1(\mu)-\mu N},
\label{inverse}
\end{align}
looks very similar to the expression of the Airy function
\eqref{airy}.
Hence, we are led to the grand canonical ensemble naturally.

However, note that there are also some discrepancies.
First, the grand potential defined in \eqref{J1def} is invariant under
the shift of $\mu$ by $2\pi i$, while the cubic polynomial in
\eqref{airy} is of course not.
Secondly, the integration domain is $[-\pi i,\pi i)$ in the inverse
transformation \eqref{inverse}, while it is the whole imaginary axis
for the Airy function \eqref{airy}.

These two discrepancies can be resolved simultaneously \cite{HMO2}.
Namely, to restore the $2\pi i$ shift symmetry, let us consider a
quantity $\Jn_1(\mu)$ and express the total grand potential
$\Jt_1(\mu)$ by infinite replicas of it,
\begin{align}
e^{\Jt_1(\mu)}=\sum_{n=-\infty}^\infty e^{\Jn_1(\mu+2\pi in)}.
\label{Jnaive}
\end{align}
If we use this quantity $\Jn_1(\mu)$, we can extend the integral
domain to the whole imaginary axis by substituting \eqref{Jnaive} into
\eqref{inverse} and connecting various intervals of integral domains
$[-\pi i+2\pi in,\pi i+2\pi in)$ with different $n$,
\begin{align}
Z_1(N)
=\int_{-\infty i}^{\infty i}\frac{d\mu}{2\pi i}
e^{\Jn_1(\mu)-\mu N}.
\label{inversenaive}
\end{align}
Note that this argument should be handled with care.
We have implicitly assumed the analyticity of $\Jn_1(\mu)$, though
generally it may contain branch cuts\footnote{Indeed the result of
small $k$ expansion \cite{MP} contains a branch cut in the
$e^\mu$-plane at $(-\infty ,-4 ]$.},
which invalidate the above argument of substituting complex-valued
chemical potentials into $\Jn_1(\mu)$ in \eqref{Jnaive}.
This assumption is supported by our numerical studies later in section
\ref{example}.

If we expand the total grand potential into\footnote{For suitability
and simplicity, we have changed the notation slightly from section 3
in \cite{HMO2}:
$[\widehat J(\mu)]^{\rm here}=[J(\mu)]^{\rm HMO}$,
$[J(\mu)]^{\rm here}=[J^{\rm naive}(\mu)]^{\rm HMO}$,
$[\widetilde J(\mu)]^{\rm here}=[J^{\rm osc}(\mu)]^{\rm HMO}$.}
\begin{align}
\Jt_1(\mu)=\Jn_1(\mu)+\widetilde J_1(\mu),
\label{osc}
\end{align}
we find that the $n\ne 0$ terms
\begin{align}
\widetilde J_1(\mu)
=\log{\Biggl[1+\frac{1}{e^{\Jn_1(\mu)}}
\sum_{n\neq 0}e^{\Jn_1(\mu +2\pi in)}\Biggr]},
\label{tildeJ}
\end{align}
give an oscillatory behavior \cite{HMO2}.
On the contrary, the $n=0$ term $\Jn_1(\mu)$ does not contain any
oscillations depending on $\mu$.
Even though the integration domain in \eqref{inversenaive} is
different from the original inverse transformation \eqref{inverse},
the extra oscillation $\widetilde J_1(\mu)$ in \eqref{tildeJ} is
completely determined by the quantity $\Jn_1(\mu)$.
Hence, we consider the $n=0$ term $\Jn_1(\mu)$ as a naive but primary
term while regard the extra $n\ne 0$ terms contributing to the
oscillatory behavior $\widetilde J_1(\mu)$ as subsidiary. 
Note that if we just wanted to restore the shift symmetry
$\mu\sim\mu+2\pi i$, at the first sight, we could define the naive
term with $n\ne 0$ as well, though this is not the case.
The $n=0$ term is characterized by the property that it does not
contain any complex phases depending on $\mu$.

The relation to the statistical mechanics was observed and fully
incardinated in \cite{MP} by identifying the partition function of the
ABJM theory as that of an ideal Fermi gas system and expressing the
grand potential in terms of the Fredholm determinant using the
density matrix $\rho_1$ of the Fermi gas system,
\begin{align}
e^{\Jt_1(\mu)}=\det(1+e^{\mu}\rho_1).
\end{align}
Then, combining the results from the topological string theory
\cite{MPtop,DMP1,DMP2,FHM,HMO2,HMMO}, the 't Hooft expansion
\cite{DMP1,DMP2,FHM}, the WKB expansion of the Fermi gas system
\cite{MP,CM}, numerical studies \cite{KEK,HMO2,HMO3} and the infinite
divergence cancellation between the worldsheet instantons and the
membrane instantons \cite{HMO2,CM,HMO3,HMMO}, we finally end up with
an exact expression, where $\Jn_1(\mu)$ consists of the perturbative
part $J_1^{\rm pert}(\mu) $ and the non-perturbative part
$J_1^{\rm np}(\mu)$,
\begin{align}
\Jn_1(\mu)=J_1^{\rm pert}(\mu)+J_1^{\rm np}(\mu),
\label{pertnp}
\end{align}
where each part is given explicitly by
\begin{align}
J_1^{\rm pert}(\mu)=\frac{C}{3}\mu^3+B\mu+A,\quad
J_1^{\rm np}(\mu)=
\sum_{\begin{subarray}{c}\ell,m=0\\(\ell,m)\ne(0,0)\end{subarray}}^\infty
f_{\ell,m}(\mu)\exp\biggl[-\biggl(2\ell+\frac{4m}{k}\biggr)\mu\biggr].
\label{JABJM}
\end{align}
The perturbative coefficients $C,B$ and $A$ are constants depending
only on $k$, while the non-perturbative coefficient $f_{\ell,m}(\mu)$
is a polynomial of $\mu$, whose explicit form can be found, for
example, in \cite{HMMO}.
The exponentially suppressed corrections $e^{-2\mu}$ and
$e^{-\frac{4\mu}{k}}$ correspond to D2-branes wrapping a Lagrangian
submanifold $\mathbb{RP}^3$ in $\mathbb{CP}^3$ and fundamental strings
wrapping a holomorphic cycle $\mathbb{CP}^1$ in it, respectively
\cite{WSinst,DMP2,MP}, where $\mathbb{CP}^3$ is obtained from
$S^7/\mathbb{Z}_k$ in the type IIA string theory limit $k\to\infty$.

Let us summarize our main result in this paper.
Here we study the partition function $Z_r(N)$ of the orbifold ABJM
theory on $S^3$, which, as explained above, is realized by the necklace
quiver $[U(N)\times U(N)]^r$ with the alternating levels $k$ and $-k$
and is expected to be dual to M-theory on
$AdS_4\times S^7/(\mathbb{Z}_{r}\times\mathbb{Z}_{kr})$.
Similarly to the ABJM matrix model, let us introduce the grand
potential
\begin{align}
e^{\Jt_r(\mu)}=\sum_{N=0}^\infty e^{\mu N}Z_r(N).
\end{align}
Again, in order to preserve the $2\pi i$ periodicity of $\mu$, we
expect the grand potential $\Jt_r(\mu)$ to be expressed as
\begin{align}
e^{\Jt_r(\mu)}=\sum_{n=-\infty}^\infty e^{\Jn_r(\mu+2\pi in)},
\label{JtJn}
\end{align}
and we shall concentrate on the primary non-oscillatory term
$\Jn_r(\mu)$.
As in the ABJM case, we can rewrite this as the Fredholm determinant
\cite{MP},
\begin{align}
e^{\Jt_r(\mu)}=\det(1+e^\mu\rho_r),
\label{rhopower}
\end{align}
with a density matrix $\rho_r$.
As we will see later in section \ref{secfermigas}, one can show that
$\rho_r$ is given by the $r$-th power of the ABJM density matrix
$\rho_1$, namely,
\begin{align}
\rho_r=\rho_1^r.
\label{rho_relation}
\end{align}
This structure leads us to the following decomposition
\begin{align}
e^{\Jt_r(\mu)}
=\prod_{j=-\frac{r-1}{2}}^{\frac{r-1}{2}}
e^{\Jt_1(\frac{\mu+2\pi ij}{r})},
\label{Jrelation}
\end{align}
where the index $j$ in the product runs with step $1$.
Thus, in terms of the primary grand potentials in \eqref{Jnaive} and
\eqref{JtJn}, the relation \eqref{Jrelation} is translated into
\begin{align}
\sum_{n=-\infty}^\infty e^{\Jn_r(\mu +2\pi in)}
=\prod_{j=-\frac{r-1}{2}}^{\frac{r-1}{2}}\sum_{n_j=-\infty}^\infty
e^{\Jn_1(\frac{\mu+2\pi ij}{r}+2\pi in_j)}.
\label{condition}
\end{align}
We would like to extract $\Jn_r(\mu)$ out of this expression.
In section \ref{derivation}, we will prove that the explicit form of
$\Jn_r(\mu)$ is given by
\begin{align}
e^{\Jn_r(\mu)}
=\sum_{\sum_jn_j=0}\prod_{j=-\frac{r-1}{2}}^{\frac{r-1}{2}}
e^{\Jn_1(\frac{\mu+2\pi ij}{r}+2\pi in_j)},
\label{repetitive}
\end{align}
where the summation symbol denotes that the summation of $r$ variables
$\bigl\{n_j\bigr\}_{j=-\frac{r-1}{2}}^{\frac{r-1}{2}}$ over all
integers is performed with the constraint $\sum_jn_j=0$.
In the original ABJM theory, we have noticed that the $n\ne 0$
replicas play only subsidiary roles.
This identity shows that the subsidiary oscillatory replicas of the ABJM
grand potential also contribute to the non-oscillatory primary term of
$\Jt_r(\mu)$, as long as the combination of the replicas satisfies the
condition $\sum_jn_j=0$.

Let us draw the physical implications of this result in section
\ref{phys}.
After substituting the ABJM grand potential \eqref{JABJM} into the
grand potential of the orbifold theory \eqref{repetitive}, we find
that the perturbative part of the grand potential $\Jn_r(\mu)$ is
given by\footnote{It was found in \cite{MP} that the grand potential
of a general ${\mathcal N}\ge 3$ necklace quiver Chern-Simons-matter
theory is given by a cubic polynomial.
Our result \eqref{Jrpert} agrees with this general claim.}
\begin{align}
J_{r}^{\rm pert}(\mu)=\frac{C}{3r^2}\mu^3
+\biggl(B-\frac{\pi^2C(r^2-1)}{3r^2}\biggr)\mu+rA,
\label{Jrpert}
\end{align}
while for the non-perturbative corrections the grand potential
$\Jn_r(\mu)$ contains the following three types of non-perturbative
instanton effects,
\begin{align}
\exp\biggl(-\frac{2\mu}{r}\biggr),\quad
\exp\biggl(-\frac{4\mu}{kr}\biggr),\quad
\exp\biggl(-\frac{4\mu}{kr^2}\biggr).
\label{instanton}
\end{align}
It is surprising to find that the last non-perturbative term in
\eqref{instanton} originally comes from the perturbative part,
\begin{align}
\exp\biggl(-\frac{2\pi^2C\mu}{r^2}\biggr),
\label{newinstanton}
\end{align}
which reduces to the last one in \eqref{instanton} after we plug in
the perturbative coefficient $C=2/(\pi^2k)$ for the ABJM matrix
model.

Let us note that, in the language of the canonical partition function, 
the perturbative part is given by
\begin{align}
Z_r^{\rm pert}(N)=\biggl(\frac{C}{r^2}\biggr)^{-1/3}e^{rA}
\Ai\biggl[\biggl(\frac{C}{r^2}\biggr)^{-1/3}
\biggl(N-B+\frac{\pi^2C(r^2-1)}{3r^2}\biggr)\biggr],
\label{Zrpert}
\end{align}
while the non-perturbative effects are described by
\begin{align}
\exp\Bigl(-\pi\sqrt{2kN}\Bigr),\quad
\exp\biggl(-2\pi\sqrt{\frac{2N}{k}}\biggr),\quad
\exp\biggl(-\frac{2\pi}{r}\sqrt{\frac{2N}{k}}\biggr),
\label{Zinst}
\end{align}
in the large $N$ limit.
As we will see later, both the perturbative part and the
non-perturbative effects match well with the gravity dual.
Especially, these non-perturbative instanton corrections correspond to
D2-branes wrapping $\mathbb{RP}^3/\mathbb{Z}_r$, fundamental strings
wrapping $\mathbb{CP}^1$ and $\mathbb{CP}^1/\mathbb{Z}_r$,
respectively.
Note that the last instanton effect arises from a new cycle compared
with the original ABJM theory.

\begin{figure}[tb]
\begin{center}
\includegraphics[scale=0.4]{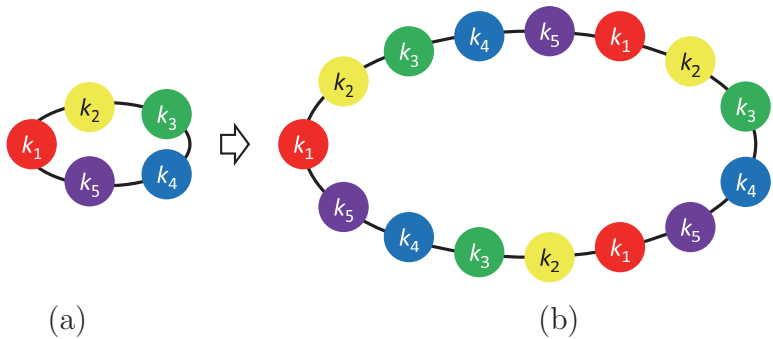}\\
\hspace{-2cm}(a)\hspace{6cm}(b)
\caption{A general quiver Chern-Simons-matter theory (b) constructed
repetitively from a more fundamental one (a) with levels satisfying
$\sum_ik_i=0$.}
\label{necklace}
\end{center}
\end{figure}

Although we start with the ABJM theory and consider its orbifold theory,
we stress that our computation is applicable to a more general setup.
Namely, even if we start with a more general $\mathcal{N}=3$ necklace
quiver Chern-Simons-matter theory having the same expression of the
grand potential \eqref{JABJM} with different coefficients and consider
its cousin with repetitive levels, the expression for the grand
potential \eqref{repetitive}, the perturbative sum \eqref{Jrpert} and
the consequent instanton effect \eqref{newinstanton} are still valid.
For example, if we have the grand potential for the necklace quiver
$\widehat A_2$ with levels $(k_1,k_2,k_3)$ ($k_1+k_2+k_3=0$), we can
easily find the grand potential for the necklace quiver
$\widehat A_{3r-1}$ with the repetitive levels
$(k_1,k_2,k_3,k_1,k_2,\cdots,k_3)$.
See figure \ref{necklace} for another case of $r=3$, constructed out
of the $\widehat A_4$ quiver Chern-Simons-matter theory.

In the following section, we briefly review the gravity dual and
discuss various possible instanton effects from it.
Then, we explain the relation \eqref{rho_relation} of the density
matrices between the orbifold ABJM theory and the original theory in
section \ref{secfermigas}, and show the relation \eqref{repetitive}
between the grand potentials in section \ref{derivation}.
In section \ref{phys} we proceed to study the physical implications.
After presenting a few examples of our result in section
\ref{example}, we conclude with some further directions in the final
section.

\section{Gravity dual}\label{gravitydual}

Before starting our study of the partition function of the 
${\mathcal N}=4$ necklace quiver Chern-Simons-matter theory with the
alternating levels, we shall first review its gravity dual \cite{ABJM}
in this section.
We also argue that we expect three types of instanton effects from the
gravity dual.
It has been expected that this ${\mathcal N}=4$ theory describes the
low-energy effective theory of M2-branes on
$\mathbb{C}^4\ni(z_1,z_2,z_3,z_4)$ divided by the following three
orbifold actions \cite{BKKS,IK,TY},
\begin{align}
\phi_A&:(z_1,z_2,z_3,z_4)
\mapsto(e^{\frac{2\pi i}{r}}z_1,e^{-\frac{2\pi i}{r}}z_2,z_3,z_4),
\nonumber\\
\phi_B&:(z_1,z_2,z_3,z_4)
\mapsto(z_1,z_2,e^{\frac{2\pi i}{r}}z_3,e^{-\frac{2\pi i}{r}}z_4),
\nonumber\\
\phi_C&:(z_1,z_2,z_3,z_4)
\mapsto(e^{\frac{2\pi i}{kr}}z_1,e^{-\frac{2\pi i}{kr}}z_2,
e^{\frac{2\pi i}{kr}}z_3,e^{-\frac{2\pi i}{kr}}z_4).
\end{align}
Since $\phi_B$ is not an independent action due to the relation
$\phi_A\phi_B=\phi_C^k$, the moduli space is
${\mathbb C}^4/({\mathbb Z}_r\times{\mathbb Z}_{kr})$.
Thus the gravity dual background of this theory is described by
\begin{align}
ds_{11}^2 =\frac{R^2}{4}ds^2_{AdS_4}
+R^2ds^2_{S^7 /(\mathbb{Z}_r\times\mathbb{Z}_{kr})},
\end{align}
where the radius $R$ is given by
\begin{align}
R=(32\pi^2kr^2N)^{\frac{1}{6}}l_p,
\end{align}
with the Planck length $l_p$.
If we identify the direction of M-theory circle $\varphi$ with that
orbifolded by ${\mathbb Z}_{kr}$,
\begin{align}
ds_{S^7 /(\mathbb{Z}_r\times\mathbb{Z}_{kr})}^2
=\frac{1}{(kr)^2}(d\varphi+\cdots)^2+ds_{\mathbb{CP}^3/\mathbb{Z}_r}^2,
\end{align}
the radius of the M-theory circle in the unit of the Planck length
$l_p$, the radius of the covering space $\mathbb{CP}^3$ in the unit of
the type IIA string length $l_s$ and the string coupling are given
respectively by
\begin{align}
\frac{R_{11}}{l_p}=\frac{R}{krl_p}
=\left(\frac{32\pi^2 N}{k^5r^4}\right)^{\frac{1}{6}},\quad
\frac{R_{\mathbb{CP}^3}}{l_s}=\frac{R}{l_s}
=\left(\frac{32\pi^2N}{k}\right)^{\frac{1}{4}},\quad
g_{s}^2=\left(\frac{32\pi^2 N}{k^5r^4}\right)^{\frac{1}{2}}.
\end{align}
Therefore, the eleven-dimensional supergravity picture is valid for
$k^5r^4\ll N$, while the type IIA supergravity is good for
$k\ll N\ll k^5r^4$.

In the eleven-dimensional supergravity on $AdS_4\times X^7$ with the
boundary $S^3$, the free energy obeys the famous $N^{3/2}$-law
\cite{KT} given by (see e.g. \cite{Mlecture} for a derivation)
\begin{align}
\log Z_{\rm supergravity}=-\sqrt{\frac{2\pi^6}{27\vol(X^7)}}N^{3/2},
\end{align} 
which relates the free energy of the current theory to that of the
ABJM theory
\begin{align}
\log Z_{S^7/(\mathbb{Z}_r\times\mathbb{Z}_{kr})}
=r\log Z_{S^7/\mathbb{Z}_{k}}.
\label{gravpert}
\end{align}

There are several non-trivial cycles in this geometry
$\mathbb{CP}^3/\mathbb{Z}_r$.
If we consider the subspace $z_1=z_2=0$, we find that the only
independent action is $\phi_C$.
Since the direction of $\mathbb{Z}_{kr}$ is identified as the M-theory
circle, we do not have any further orbifold actions.
Hence, the subspace is a holomorphic cycle $\mathbb{CP}^1$.
If we consider the subspace $z_1=z_3=0$, both of the actions $\phi_A$
and $\phi_C$ remain and we should consider the subspace as
$\mathbb{CP}^1/\mathbb{Z}_r$.
Similarly, we have a Lagrangian submanifold
$\mathbb{RP}^3/\mathbb{Z}_r$.

Since D2-branes can wrap $\mathbb{RP}^3/{\mathbb Z}_r$, we expect the
D2-brane instanton effects,
\begin{align}
\exp\Bigl(-T_{\rm D2}{\rm Vol}(\mathbb{RP}^3/{\mathbb Z}_r)\Bigr)
=\exp\biggl(-\frac{1}{(2\pi)^2l_s^3g_s}
\cdot\frac{\pi^2R_{\mathbb{CP}^3}^3}{r}\biggr)
=\exp\Bigl(-\pi\sqrt{2kN}\Bigr).
\label{D2inst}
\end{align}
Also, fundamental strings can wrap $\mathbb{CP}^1$ or
$\mathbb{CP}^1/{\mathbb Z}_r$ in this geometry and we expect two kinds
of worldsheet instanton effects given by
\begin{align}
&\exp\Bigl(-T_{\rm F1}{\rm Vol}(\mathbb{CP}^1)\Bigr)
=\exp\biggl(-\frac{1}{2\pi l_s^2}\cdot\pi R_{\mathbb{CP}^3}^2\biggr)
=\exp\biggl(-2\pi\sqrt{\frac{2N}{k}}\biggr),\nonumber\\
&\exp\Bigl(-T_{\rm F1}{\rm Vol}(\mathbb{CP}^1/\mathbb{Z}_r)\Bigr)
=\exp\biggl(-\frac{2\pi}{r}\sqrt{\frac{2N}{k}}\biggr).
\label{WSinst}
\end{align}
In section \ref{phys}, we shall reproduce these expected results of
the perturbative part \eqref{gravpert} and the non-perturbative part
\eqref{D2inst} and \eqref{WSinst}.
Before that in the subsequent sections, we shall first justify
\eqref{rho_relation} and \eqref{repetitive}.
Since the gravity dual of this ${\mathcal N}=4$ theory is the orbifold
of the ABJM theory, hereafter we shall call the corresponding quiver
Chern-Simons-matter theory the orbifold ABJM theory.

\section{Orbifold ABJM theory as a Fermi gas}\label{secfermigas}
In this section we shall show the relation between the density matrix
of the orbifold ABJM theory and that of the original theory
\eqref{rho_relation}, which enables us to express the total grand
potential of the orbifold ABJM theory in terms of that of the original
theory \eqref{Jrelation}.
By using the localization method \cite{KWY,J,HHL}, the partition
function of the orbifold ABJM theory on $S^3$ is given by
($\mu_i^{(r+1)}=\mu_i^{(1)}$)
\begin{align}
Z_r(N)=\frac{1}{(N!)^{2r}}
\int\prod_{a=1}^r\prod_{i=1}^ND\mu_i^{(a)}D\nu_i^{(a)}
\prod_{a=1}^r\frac{
\prod_{i<j}\bigl(2\sinh\frac{\mu_i^{(a)}-\mu_j^{(a)}}{2}\bigr)^2
\prod_{i<j}\bigl(2\sinh\frac{\nu_i^{(a)}-\nu_j^{(a)}}{2}\bigr)^2}
{\prod_{i,j}2\cosh\frac{\mu_i^{(a+1)}-\nu_j^{(a)}}{2}\cdot
2\cosh\frac{\nu_j^{(a)}-\mu_i^{(a)}}{2}},
\end{align}
where we have introduced the same notation as in \cite{HHMO,MM},
\begin{align}
D\mu^{(a)}_i=\frac{d\mu^{(a)}_i}{2\pi}e^{\frac{ik}{4\pi}(\mu_i^{(a)})^2},\quad
D\nu^{(a)}_i=\frac{d\nu^{(a)}_i}{2\pi}e^{-\frac{ik}{4\pi}(\nu_i^{(a)})^2}.
\label{DmuDnu}
\end{align}
Let us rewrite this into the Fermi gas formalism as in the ABJM case
\cite{MP,HHMO}.
Using the Cauchy determinant formula
\begin{align}
\frac{\prod_{i<j}2\sinh\frac{x_i-x_j}{2}
\cdot 2\sinh\frac{y_i-y_j}{2}}{\prod_{i,j}2\cosh\frac{x_i-y_j}{2}}
=\det{}\!_{i,j}\frac{1}{2\cosh\frac{x_i-y_j}{2}},
\end{align}
we find that the partition function can be rewritten into
\begin{align}
Z_r(N)=\frac{1}{(N!)^{2r}}
\int\prod_{a=1}^r\prod_{i=1}^ND\mu_i^{(a)}D\nu_i^{(a)}
\prod_{a=1}^r\det{}\!_{i,j}\frac{1}{2\cosh\frac{\mu_i^{(a+1)}-\nu_j^{(a)}}{2}}
\det{}\!_{j,i}\frac{1}{2\cosh\frac{\nu_j^{(a)}-\mu_i^{(a)}}{2}}.
\end{align}
If we expand the determinants and trivialize the permutations except
the one over $\mu_i^{(1)}$, we arrive at the following ideal Fermi gas
representation
\begin{align}
Z_r(N)&=\frac{1}{N!}
\int\prod_{i=1}^ND\mu_i^{(1)}
\sum_{\sigma\in S_N}(-1)^\sigma\prod_{i}\rho_r (\mu_{\sigma(i)}^{(1)},\mu_i^{(1)}),
\end{align}
where $\rho_r$ is the density matrix defined by
\begin{align}
\rho_r(\mu_{j}^{(1)},\mu_i^{(1)})
&=\int\prod_{a=2}^rD\mu^{(a)}\prod_{a=1}^rD\nu^{(a)}
\prod_{a=1}^r
\frac{1}{2\cosh\frac{\mu^{(a+1)}-\nu^{(a)}}{2}}
\frac{1}{2\cosh\frac{\nu^{(a)}-\mu^{(a)}}{2}}
\bigg|_{\begin{subarray}{c}\mu^{(r+1)}\to\mu^{(1)}_j\\\mu^{(1)}\to\mu^{(1)}_i\end{subarray}}.
\label{fermigas}
\end{align}
Then, it is more convenient to move to the grand canonical formalism,
\begin{align}
e^{\Jt_r(\mu)}=\sum_{N=0}^\infty Z_r(N)e^{\mu N}=\det(1+e^\mu\rho_r).
\end{align}
In the ABJM case ($r=1$), the density matrix \eqref{fermigas} is
reduced to
\begin{align}
\rho_1(\mu_{j},\mu_i)
=\int D\nu\frac{1}{2\cosh\frac{\mu_{j}-\nu}{2}}
\frac{1}{2\cosh\frac{\nu-\mu_i}{2}}.
\end{align}
If we define the matrix multiplication among $\rho_1$'s with $D\mu$
\eqref{DmuDnu}, then we easily see that two density matrices are
related by
\begin{align}
\rho_r=\rho_1^r.
\label{rhor1}
\end{align}
Hence in the present case, the total grand potential $\Jt_r(\mu)$ can
be rewritten as
\begin{align}
e^{\Jt_r(\mu)}=\det(1+e^\mu\rho_1^r)
=\prod_{j=-\frac{r-1}{2}}^{\frac{r-1}{2}}
\det(1+e^{\frac{2\pi ij}{r}}e^{\frac{\mu}{r}}\rho_1)
=\prod_{j=-\frac{r-1}{2}}^{\frac{r-1}{2}}e^{\Jt_1(\frac{\mu+2\pi ij}{r})}.
\label{total}
\end{align}

\section{Derivation of the grand potential}\label{derivation}
In this section we shall justify our expression of the grand potential
\eqref{repetitive}.
Namely, we shall prove that
\begin{itemize}
\item
after summing over the replicas, the total grand potential of the
orbifold theory reproduces the product of the total grand potential of
the original theory \eqref{total}, i.e., the grand potential of the
orbifold theory \eqref{repetitive} satisfies the relation
\eqref{condition}, and
\item
the $\sum_jn_j=0$ term does not contain oscillatory behaviors.
\end{itemize}
Namely, besides the condition \eqref{condition}, as we have noted
below \eqref{tildeJ}, the property that characterizes the naive
primary term is that it does not contain any oscillatory behavior in
$\mu$.
Hence, we should also confirm this property.

We shall prove these two facts in the following two subsections.

\subsection{Summing over replicas}
Let us show that \eqref{repetitive} satisfies the relation
\eqref{condition}, namely,
\begin{align}
\sum_{n=-\infty}^\infty\Biggl[\sum_{\sum_jn_j=0}\prod_{j=-\frac{r-1}{2}}^{\frac{r-1}{2}}
e^{\Jn_1(\frac{\mu+2\pi ij}{r}+2\pi in_j)}\Big|_{\mu\to\mu+2\pi in}\Biggr]
=\prod_{j=-\frac{r-1}{2}}^{\frac{r-1}{2}}\sum_{n_j=-\infty}^\infty
e^{\Jn_1(\frac{\mu+2\pi ij}{r}+2\pi in_j)}.
\label{problem}
\end{align}
For this purpose first let us redefine the variables in the summation
or the product as
\begin{align}
j'\equiv j+n\;(\mbox{mod $r$}),\quad
n'_j=n_j+\frac{j+n-j'}{r},
\label{redef}
\end{align}
such that $j'$ runs over the same range as $j$, namely from $-(r-1)/2$
to $(r-1)/2$ with step 1.
Then, we find that the argument of $\Jn_1$ becomes
\begin{align}
\Jn_1\biggl(\frac{\mu+2\pi ij}{r}+2\pi in_j\biggr)\bigg|_{\mu\to\mu+2\pi in}
=\Jn_1\biggl(\frac{\mu+2\pi ij'}{r}+2\pi in'_j\biggr).
\end{align}
Since we have only shifted $j$ by $n$ in \eqref{redef}, it is clear
that $j'$ runs over the same values as $j$ exactly once,
\begin{align}
\prod_{j=-\frac{r-1}{2}}^{\frac{r-1}{2}}\cdots
=\prod_{j'=-\frac{r-1}{2}}^{\frac{r-1}{2}}\cdots.
\end{align}
Note also from \eqref{redef} that $n'_j$ are all integers and the
constraint $\sum_jn_j=0$ is translated into $\sum n'_j=n$.
Hence the constraint in the summation is lifted,
\begin{align}
\sum_{n=-\infty}^\infty\sum_{\sum_jn_j=0}\cdots
=\sum_{n=-\infty}^\infty\sum_{\sum_jn'_j=n}\cdots
=\sum_{n'_j=-\infty}^\infty\cdots.
\end{align}
After removing the primes, this is nothing but the right-hand-side of
\eqref{problem}.

\subsection{No oscillations}

Next let us check that $\Jn_r(\mu)$ does not give any oscillatory
behavior.
Namely, we see that there is no imaginary $\mu$ dependence in the
exponents.

Let us first rewrite \eqref{repetitive} as
\begin{align}
\Jn_r(\mu)
=\sum_{j=-\frac{r-1}{2}}^{\frac{r-1}{2}}\Jn_1\biggl(\frac{\mu+2\pi ij}{r}\biggr)
+\log\Biggl[1+\sum_{\begin{subarray}{c}\sum_jn_j=0\\(\exists j)(n_j\ne0)\end{subarray}}
\prod_{j=-\frac{r-1}{2}}^{\frac{r-1}{2}}
\frac{e^{\Jn_1(\frac{\mu+2\pi ij}{r}+2\pi in_j)}}{e^{\Jn_1(\frac{\mu+2\pi ij}{r})}}
\Biggr].
\label{Jnr}
\end{align}
The first term coming from the sector with $n_j=0$ for all $j$
apparently contains no oscillatory terms.
For the exponents in the parenthesis coming from the sector with
$n_j\ne 0$ for some $j$, it is useful to introduce a polynomial
$g_{\ell,m}(\mu)$ determined explicitly by $f_{\ell,m}(\mu)$ in
\eqref{JABJM} and express the grand potential as
\begin{align}
e^{\Jn_1(\mu)}=e^{J_1^{\rm pert}(\mu)}\Biggl[1
+\sum_{\begin{subarray}{c}\ell,m=0\\(\ell,m)\ne(0,0)\end{subarray}}^\infty
g_{\ell,m}(\mu)e^{-(2\ell+\frac{4m}{k})\mu}\Biggr].
\end{align}
Then, the exponents become
\begin{align}
\prod_{j=-\frac{r-1}{2}}^{\frac{r-1}{2}}
\frac{e^{\Jn_1(\frac{\mu+2\pi ij}{r}+2\pi in_j)}}
{e^{\Jn_1(\frac{\mu+2\pi ij}{r})}}
=\Biggl[\prod_{j=-\frac{r-1}{2}}^{\frac{r-1}{2}}
\frac{e^{J^{\rm pert}_1(\frac{\mu+2\pi ij}{r}+2\pi in_j)}}
{e^{J^{\rm pert}_1(\frac{\mu+2\pi ij}{r})}}\Biggr]
\Biggl[1+\sum_{\begin{subarray}{c}\ell,m=0\\(\ell,m)\ne(0,0)\end{subarray}}^\infty
h_{\ell,m}(\mu;\{n_j\})e^{-(2\ell+\frac{4m}{k})\frac{\mu}{r}}\Biggr],
\label{Glm}
\end{align}
for a polynomial $h_{\ell,m}(\mu;\{n_j\})$, which also depends on
$n_j$.
If we substitute the perturbative part \eqref{JABJM} into the first
factor, we find
\begin{align}
&J_1^{\rm pert}\biggl(\frac{\mu+2\pi ij}{r}+2\pi in_j\biggr)
-J_1^{\rm pert}\biggl(\frac{\mu+2\pi ij}{r}\biggr)\nonumber\\
&=\frac{2\pi iCn_j}{r^2}\mu^2
-\frac{(2\pi)^2C}{r}n_j\biggl(n_j+\frac{2j}{r}\biggr)\mu
+2\pi in_j\biggl(B-\frac{(2\pi)^2C}{3}
\biggl(n_j^2+3\frac{n_jj}{r}+3\frac{j^2}{r^2}\biggr)\biggr).
\label{perttonp}
\end{align}
After summing over $j$ and using the condition $\sum_jn_j=0$, we find
the imaginary quadratic term vanishing, while the linear term is real.
Hence, we find that there is no oscillatory contribution.

\section{Physical implications}\label{phys}
In the previous section, we have justified our proposal
\eqref{repetitive}.
In the argument of no oscillations, we have fully utilized the
relation \eqref{perttonp}.
Here we shall see that actually this relation has further physical
implications on the perturbative part and the non-perturbative
instanton corrections.

\subsection{Perturbative part}
Let us look more carefully into the relation \eqref{perttonp} in the
previous section.
Since $2j$ satisfies $-(r-1)\le 2j\le r-1$, it is not difficult to
find that the coefficient of the linear term satisfies
\begin{align}
n_j\biggl(n_j+\frac{2j}{r}\biggr)\ge 0,
\end{align}
where the equality holds only when $n_j=0$.
Therefore, sectors with $n_j\ne 0$ for some $j$ always contain an
exponentially decaying factor and do not contribute to the
perturbative part of the orbifold theory.
Namely, the perturbative part has only contributions from the sector
with $n_j=0$ for all $j$,
\begin{align}
J_{r}^{\rm pert}(\mu)
=\sum_{j=-\frac{r-1}{2}}^{\frac{r-1}{2}}
J_{1}^{\rm pert}\biggl(\frac{\mu+2\pi ij}{r}\biggr).
\end{align}
After plugging in the expression \eqref{JABJM}, we arrive at the
expressions for $J_{r}^{\rm pert}(\mu)$ \eqref{Jrpert} and
$Z_r^{\rm pert}(N)$ \eqref{Zrpert}.
In the large $N$ limit with $k$ and $r$ fixed, the partition function
$Z_r^{\rm pert}(N)$ is expanded as
\begin{align}
\log Z_r^{\rm pert}(N)=-\frac{2}{3}rC^{-1/2}N^{3/2}
+rC^{-1/2}\biggl(B-\frac{\pi^2C(r^2-1)}{3r^2}\biggr)N^{1/2}
-\frac{1}{4}\log N+{\mathcal O}(1).
\end{align}
The first term reproduces the result \eqref{gravpert} of the classical
eleven-dimensional supergravity.
The logarithmic behavior in the third term also agrees with the 1-loop
supergravity analysis in \cite{BGMS}.

\subsection{Instanton corrections}
As we have seen in section \ref{gravitydual}, we expect the three
kinds of the instanton effects \eqref{D2inst} and \eqref{WSinst} from
the gravity dual.
Here we discuss that these instanton effects can be naturally understood 
as exponentially suppressed corrections in the grand potential $J_r (\mu)$.

First it is obvious that we have the following two corrections 
\begin{align}
\exp\biggl(-\frac{2\mu}{r}\biggr),\quad
\exp\biggl(-\frac{4\mu}{kr}\biggr) ,
\label{WSr}
\end{align}
which come from substituting $\mu/r+\cdots$ into $J_1(\mu)$ as in
\eqref{Jnr}.
Besides it, from the linear term in \eqref{perttonp} we find another
kind of exponentially suppressed correction
\begin{align}
\exp\biggl(-n\frac{2\pi^2C\mu}{r^2}\biggr),
\label{newC}
\end{align}
with the positive integer $n$ given by
\begin{align}
n=2\sum_{j=-\frac{r-1}{2}}^{\frac{r-1}{2}}n_j(n_jr+2j).
\label{selection}
\end{align}
After plugging in the ABJM value $C=2/\pi^2k$ and setting the
instanton number $n$ to be $1$, this becomes
\begin{align}
\exp\biggl(-\frac{4\mu}{kr^2}\biggr).
\label{new}
\end{align}
Using \eqref{Glm}, we find that the sectors with $n_j\ne 0$ for some
$j$ can be expressed as
\begin{align}
\sum_{\begin{subarray}{c}\sum_jn_j=0\\(\exists j)(n_j\ne0)\end{subarray}}
\prod_{j=-\frac{r-1}{2}}^{\frac{r-1}{2}}
\frac{e^{\Jn_1(\frac{\mu+2\pi ij}{r}+2\pi in_j)}}{e^{\Jn_1(\frac{\mu+2\pi ij}{r})}}
=\sum_{\begin{subarray}{c}\ell,m,n=0\\n\ne 0\end{subarray}}^\infty
g_{\ell,m,n}(\mu)
e^{-(\frac{2\ell}{r}+\frac{4m}{kr}+\frac{4n}{kr^2})\mu},
\end{align}
generally with a polynomial $g_{\ell,m,n}(\mu)$.
Thus, finally we conclude that the non-perturbative part of
$\Jn_r(\mu)$ takes the following form
\begin{align}
J^{\rm np}_r(\mu)
=\sum_{\begin{subarray}{c}\ell,m,n=0\\(\ell,m,n)\ne(0,0,0)\end{subarray}}^\infty
f_{\ell,m,n}(\mu)
\exp\biggl[-\biggl(\frac{2\ell}{r}+\frac{4m}{kr}
+\frac{4n}{kr^2}\biggr)\mu\biggr],
\end{align}
with a polynomial $f_{\ell,m,n}(\mu)$.

For comparison with our result in \eqref{D2inst} and \eqref{WSinst},
let us return to the canonical partition function.
In the large $N$ limit, the integration over the chemical potential
$\mu$ is dominated by the saddle point $\mu_\ast$ related to $N$ by
\begin{align}
\frac{\partial J_r(\mu)}{\partial\mu}\bigg|
_{\mu=\mu_\ast} =N,
\quad\mu_\ast=\pi r\sqrt{\frac{kN}{2}}.
\end{align}
Hence, the exponentially suppressed corrections \eqref{WSr} and
\eqref{new} on the gauge theory side are translated to \eqref{Zinst}
\begin{align}
\exp\Bigl(-\pi\sqrt{2kN}\Bigr),\quad
\exp\biggl(-2\pi\sqrt{\frac{2N}{k}}\biggr),\quad
\exp\biggl(-\frac{2\pi}{r}\sqrt{\frac{2N}{k}}\biggr),
\nonumber
\end{align}
which are exactly the same instanton effects found from the gravity
side in \eqref{D2inst} and \eqref{WSinst}.
For finite $N$, these instanton effects contribute to $Z_r(N)$ by a
superposition of
\begin{align}
\Ai\biggl[\biggl(\frac{C}{r^2}\biggr)^{-1/3}
\biggl(N-B+\frac{\pi^2C(r^2-1)}{3r^2}
+\frac{2\ell}{r}+\frac{4m}{kr}+\frac{4n}{kr^2}\biggr)\biggr]
\end{align}
and their derivatives.

Note that unlike the original instanton effects, $n$ obeys an
interesting selection rule and only takes a discrete set of integers
\eqref{selection}.
For example, $n$ always has to be an even number and the smallest
instanton number is $n=4$ which comes from the combination of $n_j$
with its non-zero components given by
\begin{align}
n_{-\frac{r-1}{2}}=1,\quad n_{\frac{r-1}{2}}=-1.
\end{align}

\section{Examples}\label{example}

Let us use our formula \eqref{repetitive} to write down several terms
explicitly.
Note that once we have a general formula, all these results can be
obtained very easily from \cite{HMO2,HMO3}.
One purpose of this section is to see a rough structure, which would
be useful in the future study of more general ${\mathcal N}=3$
Chern-Simons-matter theories.
In the following we will show the non-perturbative part of the grand
potential $J^{\rm np}_{r,k}(\mu)$ for the case of $k=2$.
The reason we choose this case is because of its novel behavior,
though we can study for any $r$ and any $k$.
\begin{align}
&J_{2,2}^{\rm np}(\mu)
=\biggl[-\frac{2\mu^2+2\mu+2}{\pi^2}+2\biggr]e^{-\mu}
+\biggl[-\frac{52\mu^2+2\mu+9}{4\pi^2}+18\biggr]e^{-2\mu}
\nonumber\\&\qquad
+\biggl[-\frac{368\mu^2-304\mu/3+308/9}{3\pi^2}
+\frac{608}{3}\biggr]e^{-3\mu}
\nonumber\\&\qquad
+\biggl[-\frac{2701\mu^2-13949\mu/12+11291/48}{2\pi^2}
+2514\biggr]e^{-4\mu}+{\mathcal O}(e^{-5\mu}),
\label{J22}
\\
&J_{3,2}^{\rm np}(\mu)
=\biggl[-\frac{4(4\mu+3)}{3\sqrt{3}\pi}+\frac{16}{9}\biggr]e^{-\frac{2}{3}\mu}
+e^{-\frac{8}{9}\mu}
+\biggl[-\frac{104\mu+3}{3\sqrt{3}\pi}-\frac{104}{9}\biggr]e^{-\frac{4}{3}\mu}
\nonumber\\&\qquad
+\biggl[\frac{4(4\mu+3)}{\sqrt{3}\pi}+\frac{16}{3}\biggr]e^{-\frac{14}{9}\mu}
+{\mathcal O}(e^{-\frac{16}{9}\mu}),
\label{J32}
\\
&J_{4,2}^{\rm np}(\mu)
=\biggl[-\frac{2\mu+2}{\pi}+1\biggr]e^{-\frac{1}{2}\mu}
+\biggl[\frac{13\mu^2+\mu+9}{2\pi^2}+\frac{8(\mu+1)}{\pi}-43\biggr]e^{-\mu}
\nonumber\\&\qquad
+\biggl[\frac{32(\mu+1)^2}{\pi^2}+\frac{344\mu-376/3}{3\pi}
-\frac{152}{3}\biggr]e^{-\frac{3}{2}\mu}
\nonumber\\&\qquad
+\biggl[\frac{256(\mu+1)^3}{3\pi^3}
-\frac{2957\mu^2-10877\mu/6+14363/12}{4\pi^2}
-\frac{2720\mu+2528/3}{3\pi}+5754\biggr]e^{-2\mu}
\nonumber\\&\qquad
+{\mathcal O}(e^{-\frac{5}{2}\mu}).
\label{J42}
\end{align}
Note that for the case of $r=2$ the coefficients are very similar to
the $(k,M)=(4,1)$ case of the ABJ theory \cite{MM}, which is the
${\mathcal N}=6$ Chern-Simons-matter theory with gauge group 
$U(N)\times U(N+M)$ and levels $k$ and $-k$ \cite{ABJ,HLLLP2}.
For the case of $r=3$, the novel instanton term
$e^{-\frac{16\mu}{kr^2}}$ discussed previously does not mix with other
contributions.
For the case of $r=4$, the coefficient polynomials of instantons are
not necessarily quadratic any more.

Forgetting about the result \eqref{repetitive}, we have also checked
\eqref{J22}-\eqref{J42} numerically by the same method used in
\cite{HMO2}.
Namely, using the exact values of the partition function $Z_r(N)$ in
the orbifold ABJM theory obtained from those in the original theory
with the relation \eqref{rho_relation}, we try to find out the exact
coefficients of the primary grand potential $J_r(\mu)$ numerically
using the inverse transformation
\begin{align}
Z_r(N)=\int_{-\infty i}^{\infty i}\frac{d\mu}{2\pi i}e^{J_r(\mu)-\mu N}.
\end{align}
We find that the coefficients match well with those in
\eqref{J22}-\eqref{J42} within about 1\% errors.
Since we have less exact values (with the number divided by $r$) and
relatively milder decaying factors in the instanton effects
(especially for the $r=3$ case due to the intermediate new instanton
effect \eqref{new}), it is difficult to check with high precision to
high instanton corrections.
However, we believe that at least it is safe to claim that we have
detected the appearance of the new instanton numerically.\footnote{We
are grateful to Y.~Hatsuda for valuable discussions.}

\section{Discussions}\label{discuss}
We have studied the partition function of the orbifold ABJM theory via
the grand canonical formalism.
We have found the explicit formula \eqref{repetitive} for the grand
potential in terms of the grand potential of the ABJM theory.
It is surprising to find that the subsidiary oscillatory terms of the
original theory lead to the primary non-oscillatory term of the
orbifold theory and the perturbative part in the original theory
results in the new non-perturbative instanton effect
\eqref{newinstanton} in the orbifold theory.
We have identified this instanton effect as the string worldsheet
wrapping the cycle $\mathbb{CP}^1/{\mathbb Z}_r$ in
${\mathbb C}^4/({\mathbb Z}_{r}\times{\mathbb Z}_{kr})$.
Let us discuss some future directions.

As we have explained in section \ref{intro}, there are several methods
to study the matrix model.
We would like to detect the new instanton effect from complementary
methods.
For example, since the density matrix of the orbifold theory is
related directly to that of the ABJM theory \eqref{rho_relation}, it
is great to see how the grand potential is also derived from the exact
quantization as in \cite{KM}.
Also, we expect that the new instanton effects can be reproduced from
the genus 1 analysis in the 't Hooft limit as in \cite{DMP1}.

It has been found that the new instanton effect obeys the interesting
selection rule \eqref{selection}.
It is important to understand the origin of the selection rule both
from the direct field theoretical study and from the gravity dual
\cite{BKKS,IK,TY}.

The result that the perturbative part in the original theory results in
the non-perturbative effect in the orbifold theory \eqref{newinstanton}
may look less surprising in view of the following interpretation from
the gravity dual.\footnote{We are grateful to the referee for valuable
comments.}
Since this non-perturbative effect corresponds to the string worldsheet
wrapping the cycle $\mathbb{CP}^1/{\mathbb Z}_r$ which is not present
in the original $AdS_4\times S^7/\mathbb{Z}_k$ background, they can
only have a perturbative origin.
We hope to understand it from the field theoretical study as well.

In the context of resurgence theory, it was found \cite{B,ZJ,DU}
that the perturbative asymptotic expansion contains the information of
the non-perturbative effects.
In extended supersymmetric theories \cite{R}, even though there are no
ambiguities in the asymptotic expansion, our result still shows that
the perturbative coefficient appears in the non-perturbative effects
\eqref{newinstanton}.
It would be great to understand more extensively also from the
viewpoint of the resurgence theory.

We have often substituted the complex values of the chemical potential
$\mu$ into the primary non-oscillatory grand potential of the ABJM
theory as in \eqref{repetitive}.
However, the original expression is only literally valid for large
real chemical potential $\mu$.
This is allowed only when they are in the same Stokes sector.
Our numerical check in section \ref{example} is a crucial support for
this assumption, though we hope to study the analytical structure of
the grand potential carefully.

Our result on the orbifold ABJM theory \eqref{repetitive} has been
obtained based on the result of the ABJM theory.
We hope that there is a more direct expression with topological
invariants on a certain dual geometry of the orbifold ABJM theory.

As we have noted in section \ref{intro}, our expression of the grand
potential \eqref{repetitive} including the perturbative sum
\eqref{Jrpert} and the new instanton effect \eqref{newinstanton} are
very general.
Namely, they are applicable to any necklace quiver Chern-Simons-matter
theory as long as the quiver diagram is repetitive as in figure
\ref{necklace}.
We hope to study more general theories using these results.
For the ABJ theory \cite{ABJ,HLLLP2}, in the Fermi gas formalism of
\cite{AHS,H}, the grand potential is expressed in a similar manner.
This means that our result is also applicable to these cases, though
this fact is not so obvious from the formalism of \cite{MM}.

The meaning of various sectors with different combinations of $n_j$ is
not very clear.
In some sense, the sector with $n_j=0$ for all $j$ is similar to the
untwisted sector in the string orbifold theory, while the others are
similar to the twisted ones.
We hope to clarify the physical meaning of these sectors.

\section*{Acknowledgements}
We are grateful to S.~Hirano, Y.~Imamura, J.~K\"all\'en, M.~Mari\~no,
K.~Ohta, K.~Okuyama, J.~Park, M.~Shigemori, A.~Tanaka, M.~Unsal,
S.~Yamaguchi and especially Y.~Hatsuda for valuable discussions.
M.~H.\ would like to thank Kavli IPMU and S.~M.\ would like to thank
KEK for warm hospitality.
We would also like to thank the participants of the JSPS/RFBR
conference at Biwako for important feedbacks, where part of this work
was presented by S.~M.


\begin{thebibliography}{99}
\bibitem{ABJM}
O.~Aharony, O.~Bergman, D.~L.~Jafferis and J.~Maldacena,
``N=6 superconformal Chern-Simons-matter theories, M2-branes and their
gravity duals,''
JHEP {\bf 0810}, 091 (2008)
[arXiv:0806.1218 [hep-th]].
%%CITATION = ARXIV:0806.1218;%%
\bibitem{S}
J.~H.~Schwarz,
``Superconformal Chern-Simons theories,''
JHEP {\bf 0411}, 078 (2004)
[hep-th/0411077].
%%CITATION = HEP-TH/0411077;%%
\bibitem{ZK} 
B.~M.~Zupnik and D.~V.~Khetselius,
``Three-dimensional extended supersymmetry in the harmonic
superspace,''
Sov.\ J.\ Nucl.\ Phys.\  {\bf 47}, 730 (1988)
[Yad.\ Fiz.\  {\bf 47}, 1147 (1988)].
%%CITATION = SJNCA,47,730;%%
\bibitem{KL} 
H.~-C.~Kao and K.~-M.~Lee,
``Selfdual Chern-Simons systems with an N=3 extended supersymmetry,''
Phys.\ Rev.\ D {\bf 46}, 4691 (1992)
[hep-th/9205115].
%%CITATION = HEP-TH/9205115;%%
\bibitem{KLL}
H.~-C.~Kao, K.~-M.~Lee and T.~Lee,
``The Chern-Simons coefficient in supersymmetric Yang-Mills
Chern-Simons theories,''
Phys.\ Lett.\ B {\bf 373}, 94 (1996)
[hep-th/9506170].
%%CITATION = HEP-TH/9506170;%%
\bibitem{GW}
D.~Gaiotto and E.~Witten,
``Janus Configurations, Chern-Simons Couplings, And The theta-Angle in
N=4 Super Yang-Mills Theory,''
JHEP {\bf 1006}, 097 (2010)
[arXiv:0804.2907 [hep-th]].
%%CITATION = ARXIV:0804.2907;%%
\bibitem{HLLLP1} 
K.~Hosomichi, K.~-M.~Lee, S.~Lee, S.~Lee and J.~Park,
``N=4 Superconformal Chern-Simons Theories with Hyper and Twisted
Hyper Multiplets,''
JHEP {\bf 0807}, 091 (2008)
[arXiv:0805.3662 [hep-th]].
%%CITATION = ARXIV:0805.3662;%%
\bibitem{BKKS}
M.~Benna, I.~Klebanov, T.~Klose and M.~Smedback,
``Superconformal Chern-Simons Theories and AdS(4)/CFT(3) Correspondence,''
JHEP {\bf 0809}, 072 (2008)
[arXiv:0806.1519 [hep-th]].
%%CITATION = ARXIV:0806.1519;%%
\bibitem{IK}
Y.~Imamura and K.~Kimura,
``On the moduli space of elliptic Maxwell-Chern-Simons theories,''
Prog.\ Theor.\ Phys.\  {\bf 120}, 509 (2008)
[arXiv:0806.3727 [hep-th]].
%%CITATION = ARXIV:0806.3727;%%
\bibitem{TY}
S.~Terashima and F.~Yagi,
``Orbifolding the Membrane Action,''
JHEP {\bf 0812}, 041 (2008)
[arXiv:0807.0368 [hep-th]].
%%CITATION = ARXIV:0807.0368;%%
\bibitem{IK4}
Y.~Imamura and K.~Kimura,
``N=4 Chern-Simons theories with auxiliary vector multiplets,''
JHEP {\bf 0810}, 040 (2008)
[arXiv:0807.2144 [hep-th]].
%%CITATION = ARXIV:0807.2144;%%
\bibitem{DT} 
N.~Drukker and D.~Trancanelli,
``A Supermatrix model for N=6 super Chern-Simons-matter theory,''
JHEP {\bf 1002}, 058 (2010)
[arXiv:0912.3006 [hep-th]].
%%CITATION = ARXIV:0912.3006;%%
\bibitem{MPtop} 
M.~Marino and P.~Putrov,
``Exact Results in ABJM Theory from Topological Strings,''
JHEP {\bf 1006}, 011 (2010)
[arXiv:0912.3074 [hep-th]].
%%CITATION = ARXIV:0912.3074;%%
\bibitem{DMP1}
N.~Drukker, M.~Marino and P.~Putrov,
``From weak to strong coupling in ABJM theory,''
Commun.\ Math.\ Phys.\  {\bf 306}, 511 (2011)
[arXiv:1007.3837 [hep-th]].
%%CITATION = ARXIV:1007.3837;%%
\bibitem{HKPT}
C.~P.~Herzog, I.~R.~Klebanov, S.~S.~Pufu and T.~Tesileanu,
``Multi-Matrix Models and Tri-Sasaki Einstein Spaces,''
Phys.\ Rev.\ D {\bf 83}, 046001 (2011)
[arXiv:1011.5487 [hep-th]].
%%CITATION = ARXIV:1011.5487;%%
\bibitem{DMP2} 
N.~Drukker, M.~Marino and P.~Putrov,
``Nonperturbative aspects of ABJM theory,''
JHEP {\bf 1111}, 141 (2011)
[arXiv:1103.4844 [hep-th]].
%%CITATION = ARXIV:1103.4844;%%
\bibitem{FHM}
H.~Fuji, S.~Hirano and S.~Moriyama,
``Summing Up All Genus Free Energy of ABJM Matrix Model,''
JHEP {\bf 1108}, 001 (2011)
[arXiv:1106.4631 [hep-th]].
%%CITATION = ARXIV:1106.4631;%%
\bibitem{O} 
K.~Okuyama,
``A Note on the Partition Function of ABJM theory on $S^3$,''
Prog.\ Theor.\ Phys.\  {\bf 127}, 229 (2012)
[arXiv:1110.3555 [hep-th]].
%%CITATION = ARXIV:1110.3555;%%
\bibitem{MP}
M.~Marino and P.~Putrov,
``ABJM theory as a Fermi gas,''
J.\ Stat.\ Mech.\  {\bf 1203}, P03001 (2012)
[arXiv:1110.4066 [hep-th]].
%%CITATION = ARXIV:1110.4066;%%
\bibitem{KEK}
M.~Hanada, M.~Honda, Y.~Honma, J.~Nishimura, S.~Shiba and Y.~Yoshida,
``Numerical studies of the ABJM theory for arbitrary N at arbitrary
coupling constant,''
JHEP {\bf 1205}, 121 (2012)
[arXiv:1202.5300 [hep-th]].
%%CITATION = ARXIV:1202.5300;%%
\bibitem{KMSS} 
A.~Klemm, M.~Marino, M.~Schiereck and M.~Soroush,
``ABJM Wilson loops in the Fermi gas approach,''
arXiv:1207.0611 [hep-th].
%%CITATION = ARXIV:1207.0611;%%
\bibitem{HMO1}
Y.~Hatsuda, S.~Moriyama and K.~Okuyama,
``Exact Results on the ABJM Fermi Gas,''
JHEP {\bf 1210}, 020 (2012)
[arXiv:1207.4283 [hep-th]].
%%CITATION = ARXIV:1207.4283;%%
\bibitem{PY}
P.~Putrov and M.~Yamazaki,
``Exact ABJM Partition Function from TBA,''
Mod.\ Phys.\ Lett.\ A {\bf 27}, 1250200 (2012)
[arXiv:1207.5066 [hep-th]].
%%CITATION = ARXIV:1207.5066;%%
\bibitem{HMO2}
Y.~Hatsuda, S.~Moriyama and K.~Okuyama,
``Instanton Effects in ABJM Theory from Fermi Gas Approach,''
JHEP {\bf 1301}, 158 (2013)
[arXiv:1211.1251 [hep-th]].
%%CITATION = ARXIV:1211.1251;%%
\bibitem{CM}
F.~Calvo and M.~Marino,
``Membrane instantons from a semiclassical TBA,''
JHEP {\bf 1305}, 006 (2013)
[arXiv:1212.5118 [hep-th]].
%%CITATION = ARXIV:1212.5118;%%
\bibitem{HMO3}
Y.~Hatsuda, S.~Moriyama and K.~Okuyama,
``Instanton Bound States in ABJM Theory,''
JHEP {\bf 1305}, 054 (2013)
[arXiv:1301.5184 [hep-th]].
%%CITATION = ARXIV:1301.5184;%%
\bibitem{GKM}
A.~Grassi, J.~Kallen and M.~Marino,
``The topological open string wavefunction,''
arXiv:1304.6097 [hep-th].
%%CITATION = ARXIV:1304.6097;%%
\bibitem{HMMO}
Y.~Hatsuda, M.~Marino, S.~Moriyama and K.~Okuyama,
``Non-perturbative effects and the refined topological string,''
arXiv:1306.1734 [hep-th].
%%CITATION = ARXIV:1306.1734;%%
\bibitem{HHMO}
Y.~Hatsuda, M.~Honda, S.~Moriyama and K.~Okuyama,
``ABJM Wilson Loops in Arbitrary Representations,''
JHEP {\bf 1310}, 168 (2013)
[arXiv:1306.4297 [hep-th]].
%%CITATION = ARXIV:1306.4297;%%
\bibitem{KM}
J.~Kallen and M.~Marino,
``Instanton effects and quantum spectral curves,''
arXiv:1308.6485 [hep-th].
%%CITATION = ARXIV:1308.6485;%%
\bibitem{P}
V.~Pestun,
``Localization of gauge theory on a four-sphere and supersymmetric
Wilson loops,''
Commun.\ Math.\ Phys.\  {\bf 313}, 71 (2012)
[arXiv:0712.2824 [hep-th]].
%%CITATION = ARXIV:0712.2824;%%
\bibitem{KWY}
A.~Kapustin, B.~Willett and I.~Yaakov,
``Exact Results for Wilson Loops in Superconformal Chern-Simons
Theories with Matter,''
JHEP {\bf 1003}, 089 (2010)
[arXiv:0909.4559 [hep-th]].
%%CITATION = ARXIV:0909.4559;%%
\bibitem{J} 
D.~L.~Jafferis,
``The Exact Superconformal R-Symmetry Extremizes Z,''
JHEP {\bf 1205}, 159 (2012)
[arXiv:1012.3210 [hep-th]].
%%CITATION = ARXIV:1012.3210;%%
\bibitem{HHL}
N.~Hama, K.~Hosomichi and S.~Lee,
``Notes on SUSY Gauge Theories on Three-Sphere,''
JHEP {\bf 1103}, 127 (2011)
[arXiv:1012.3512 [hep-th]].
%%CITATION = ARXIV:1012.3512;%%
\bibitem{KWYmirror} 
A.~Kapustin, B.~Willett and I.~Yaakov,
``Nonperturbative Tests of Three-Dimensional Dualities,''
JHEP {\bf 1010}, 013 (2010)
[arXiv:1003.5694 [hep-th]].
%%CITATION = ARXIV:1003.5694;%%
\bibitem{MS} 
D.~Martelli and J.~Sparks,
``The large N limit of quiver matrix models and Sasaki-Einstein manifolds,''
Phys.\ Rev.\ D {\bf 84}, 046008 (2011)
[arXiv:1102.5289 [hep-th]].
%%CITATION = ARXIV:1102.5289;%%
\bibitem{CHH} 
S.~Cheon, H.~Kim and N.~Kim,
``Calculating the partition function of N=2 Gauge theories on $S^3$
and AdS/CFT correspondence,''
JHEP {\bf 1105}, 134 (2011)
[arXiv:1102.5565 [hep-th]].
%%CITATION = ARXIV:1102.5565;%%
\bibitem{JKPS}
D.~L.~Jafferis, I.~R.~Klebanov, S.~S.~Pufu and B.~R.~Safdi,
``Towards the F-Theorem: N=2 Field Theories on the Three-Sphere,''
JHEP {\bf 1106}, 102 (2011)
[arXiv:1103.1181 [hep-th]].
%%CITATION = ARXIV:1103.1181;%%
\bibitem{GHP}
D.~R.~Gulotta, C.~P.~Herzog and S.~S.~Pufu,
``From Necklace Quivers to the F-theorem, Operator Counting, and
T(U(N)),''
JHEP {\bf 1112}, 077 (2011)
[arXiv:1105.2817 [hep-th]].
%%CITATION = ARXIV:1105.2817;%%
\bibitem{GAH}
D.~R.~Gulotta, J.~P.~Ang and C.~P.~Herzog,
``Matrix Models for Supersymmetric Chern-Simons Theories with an ADE
Classification,''
JHEP {\bf 1201}, 132 (2012)
[arXiv:1111.1744 [hep-th]].
%%CITATION = ARXIV:1111.1744;%%
\bibitem{GHN}
D.~R.~Gulotta, C.~P.~Herzog and T.~Nishioka,
``The ABCDEF's of Matrix Models for Supersymmetric Chern-Simons Theories,''
JHEP {\bf 1204}, 138 (2012)
[arXiv:1201.6360 [hep-th]].
%%CITATION = ARXIV:1201.6360;%%
\bibitem{interacting} 
M.~Marino and P.~Putrov,
``Interacting fermions and N=2 Chern-Simons-matter theories,''
JHEP {\bf 1311}, 199 (2013)
[arXiv:1206.6346 [hep-th]].
%%CITATION = ARXIV:1206.6346;%%
\bibitem{MePu}
M.~Mezei and S.~S.~Pufu,
``Three-sphere free energy for classical gauge groups,''
arXiv:1312.0920 [hep-th].
\bibitem{GM}
A.~Grassi and M.~Marino,
``M-theoretic matrix models,''
arXiv:1403.4276 [hep-th].
%%CITATION = ARXIV:1403.4276;%%
\bibitem{SMP}
R.~C.~Santamaria, M.~Marino and P.~Putrov,
``Unquenched flavor and tropical geometry in strongly coupled
Chern-Simons-matter theories,''
JHEP {\bf 1110}, 139 (2011)
[arXiv:1011.6281 [hep-th]].
%%CITATION = ARXIV:1011.6281;%%
\bibitem{Suyama1} 
T.~Suyama,
``Eigenvalue Distributions in Matrix Models for Chern-Simons-matter
Theories,''
Nucl.\ Phys.\ B {\bf 856}, 497 (2012)
[arXiv:1106.3147 [hep-th]].
%%CITATION = ARXIV:1106.3147;%%
\bibitem{Suyama2} 
T.~Suyama,
``On Large N Solution of N=3 Chern-Simons-adjoint Theories,''
Nucl.\ Phys.\ B {\bf 867}, 887 (2013)
[arXiv:1208.2096 [hep-th]].
%%CITATION = ARXIV:1208.2096;%%
\bibitem{Suyama3} 
T.~Suyama,
``A Systematic Study on Matrix Models for Chern-Simons-matter Theories,''
Nucl.\  Phys.\ B {\bf  874}, 528 (2013)
[arXiv:1304.7831 [hep-th]].
%%CITATION = ARXIV:1304.7831;%%
\bibitem{WSinst} 
A.~Cagnazzo, D.~Sorokin and L.~Wulff,
``String instanton in AdS(4) x CP**3,''
JHEP {\bf 1005}, 009 (2010)
[arXiv:0911.5228 [hep-th]].
%%CITATION = ARXIV:0911.5228;%%
\bibitem{KT}
I.~R.~Klebanov and A.~A.~Tseytlin,
``Entropy of near extremal black $p$-branes,''
Nucl.\ Phys.\ B {\bf 475}, 164 (1996)
[hep-th/9604089].
%%CITATION = HEP-TH/9604089;%%
\bibitem{Mlecture} 
M.~Marino,
``Lectures on localization and matrix models in supersymmetric
Chern-Simons-matter theories,''
J.\ Phys.\ A {\bf 44}, 463001 (2011)
[arXiv:1104.0783 [hep-th]].
%%CITATION = ARXIV:1104.0783;%%
\bibitem{MM}
S.~Matsumoto and S.~Moriyama,
``ABJ Fractional Brane from ABJM Wilson Loop,''
JHEP {\bf 1403}, 079 (2014)
arXiv:1310.8051 [hep-th].
%%CITATION = ARXIV:1310.8051;%%
\bibitem{BGMS}
S.~Bhattacharyya, A.~Grassi, M.~Marino and A.~Sen,
``A One-Loop Test of Quantum Supergravity,''
Class.\ Quant.\ Grav.\ {\bf 31}, 015012 (2014)
[arXiv:1210.6057 [hep-th]].
%%CITATION = ARXIV:1210.6057;%%
\bibitem{ABJ}
O.~Aharony, O.~Bergman and D.~L.~Jafferis,
``Fractional M2-branes,''
JHEP {\bf 0811}, 043 (2008)
[arXiv:0807.4924 [hep-th]].
%%CITATION = ARXIV:0807.4924;%%
\bibitem{HLLLP2} 
K.~Hosomichi, K.~-M.~Lee, S.~Lee, S.~Lee and J.~Park,
``N=5,6 Superconformal Chern-Simons Theories and M2-branes on Orbifolds,''
JHEP {\bf 0809}, 002 (2008)
[arXiv:0806.4977 [hep-th]].
%CITATION = ARXIV:0806.4977;%%
\bibitem{AHS} 
H.~Awata, S.~Hirano and M.~Shigemori,
``The Partition Function of ABJ Theory,''
Prog.\ Theor.\ Exp.\ Phys.\ 053B04 (2013)
[arXiv:1212.2966].
%%CITATION = ARXIV:1212.2966;%%
\bibitem{H}
M.~Honda,
``Direct derivation of "mirror" ABJ partition function,''
JHEP {\bf 1312}, 046 (2013)
[arXiv:1310.3126 [hep-th]].
%%CITATION = ARXIV:1310.3126;%%
\bibitem{B}
E.~B.~Bogomolny,
``Calculation of instanton-anti-instanton contributions in quantum
mechanics,''
Phys.\ Lett.\ {\bf B91} (1980), 431-435. 
\bibitem{ZJ}
J.~Zinn-Justin, 
``Multi-instanton contributions in quantum mechanics,''
Nucl.\ Phys.\ {\bf B192} (1981), 125-140.
\bibitem{DU}
G.~V.~Dunne and M.~Unsal,
``Uniform WKB, Multi-instantons, and Resurgent Trans-Series,''
arXiv:1401.5202 [hep-th].
%%CITATION = ARXIV:1401.5202;%%
\bibitem{R}
J.~G.~Russo,
``A Note on perturbation series in supersymmetric gauge theories,''
JHEP {\bf 1206}, 038 (2012)
[arXiv:1203.5061 [hep-th]].
%%CITATION = ARXIV:1203.5061;%%
\end{thebibliography}
\end{document}